\documentclass[%
 reprint,
 amsmath,amssymb,
 aps,
]{revtex4-2}

\usepackage{graphicx}
\usepackage{dcolumn}
\usepackage{bm}

\begin{document}

\preprint{APS/123-QED}

\title{A Highly Efficient and Pure Few-Photon Source on Chip}

\author{Zhaohui Ma}
\author{Jia-Yang Chen, Malvika Garikapati, Zhan Li, Chao Tang, Yong Meng Sua}
\author{Yu-Ping Huang}%
 \email{yhuang5@stevens.edu}
\affiliation{%
 Department of Physics, Stevens Institute of Technology, 1 Castle Point Terrace, Hoboken, New Jersey, 07030, USA
}%
\affiliation{
 Center for Quantum Science and Engineering, Stevens Institute of Technology, 1 Castle Point Terrace, Hoboken, New Jersey, 07030, USA
}%


\date{\today}

\begin{abstract}

We report on multi-photon statistics of correlated twin beams produced in a periodic poled micro-ring resonator on thin-film lithium niobate. Owing to high cavity confinement and near perfect quasi-phase matching, the photons pairs are produced efficiently in single modes at rates reaching 27 MHz per $\mu$W pump power. By using a pump laser whose pulse width impedance matches with the cavity, those photons are further created in single longitudinal modes with purity reaching 99\%, without relying on later-on filtering. With a dual-channel photon-number resolving detection system, we obtain directly the joint detection probabilities of multi-photon states up to three photons, with high coincidence to accidental contrast for each. Used as a single photon source, it gives heralded $g_H^{(2)}(0)$ around 0.04 at a single photon rate of 650 kHz on chip. The findings of our research highlight the potential of this nanophotonic platform as a promising platform for generating non-classical, few-photon states with ideal indistinguishability, for fundamental quantum optics studies and information applications.

\end{abstract}

\maketitle


\section{\label{sec:level1}Introduction}


Discrete photon-number and quantum entangled states are among the cornerstones of quantum optics and its many information processing applications. Limited by photon creation and measurement technology, most quantum applications hitherto have been designed based on the uses of their lowest-order forms: single photons or two of them in pairs. For example, quantum key distribution based on BB84 uses antibunched single photon states\cite{bennet1984quantum}, while quantum teleportation takes advantage of two-photon entanglement\cite{ bennett1993teleporting}. Lately, the emergence of photon-number resolving (PNR) capability in photon detection has open a door to a new paradigm of quantum optics, where nonclassical states containing multiple photons promise to offer significant advantages in computing and sensing. In this pursuit, encouraging progress has been made in the generation of multiphoton quantum states \cite{avenhaus2010accessing,harder2016single}, quantum interferometry using N00N states \cite{thekkadath2020quantum,qin2023unconditional}, quantum sensing using photon-number squeezing \cite{frascella2021overcoming}, and quantum computing \cite{arrazola2021quantum}.

To capitalize on the quantum benefits of multiphoton states, it is desirable to embed them in single optical modes. In bulky photon sources of spontaneous parametric downconversion (SPDC) or four-wave mixing, to meet this condition usually requires ultra-narrow band filtering or using ultra-short, broadband pump pulses \cite{harder2016single,eckstein2011highly,avenhaus2008photon}, either of which add significantly to system complexity and footprint. In contrast, nanophotonic circuits with high Q cavities can create photons intrinsically in single spatial and temporal modes of high purity. For example, a $\chi^{(3)}$ microring was shown to produce squeezed states in good single modes, albeit suffering parametric fluorescence emission into multiple cavity lines \cite{vaidya2020broadband}. 

Here, we demonstrate an on-chip $\chi^{(2)}$ source of multiphoton states in quasi-phase matched microrings of lithium niobate on insulator (LNOI). Due to subwavelength lateral confinement, the photons are created in single transverse (spatial) modes of high purity. With a high cavity Q and by using a pump laser whose pulse width impedance-matches with the cavity, those photons are further created in single longitudinal (time-frequency) modes with purity reaching 99\%, without relying on later-on filtering. Such high purity in both spatial and time-frequency modes gives rise to high indistinguishability, as desirable for many quantum computing, teleportation, and sensing applications. Aided by nearly perfect quasi phase matching through periodic poling, the photon generation efficiency is exceptional, where only microwatt pump power is required to create single, double, and triplet photon states of high correlation and at megahertz rates. Such high purity and high efficiency contribute to the device  scaling and wide deployment. Together with narrow cavity bandwidth, they suppress background noise created through, e.g., Raman scattering or fluorescence emission. On detection, we use photon-number resolving, superconducting nanowire single-photon detectors (PNR-SNSPDs) built in a parallel circuit configuration to accurately characterize the photon number statistics and time correlation of multiphoton states with picosecond resolution. Our results show high coincident to accidental ratios for photon counts in one, two, and three photon states. Finally, we show how this system can be used for heralded single-photon generation at 10 MHz clock speed. \cite{davis2022improved,stasi2022enhanced}.

\begin{figure}[ht]
  \label{fig: Device Calibration}
  \centering
\includegraphics[width=3.2in]{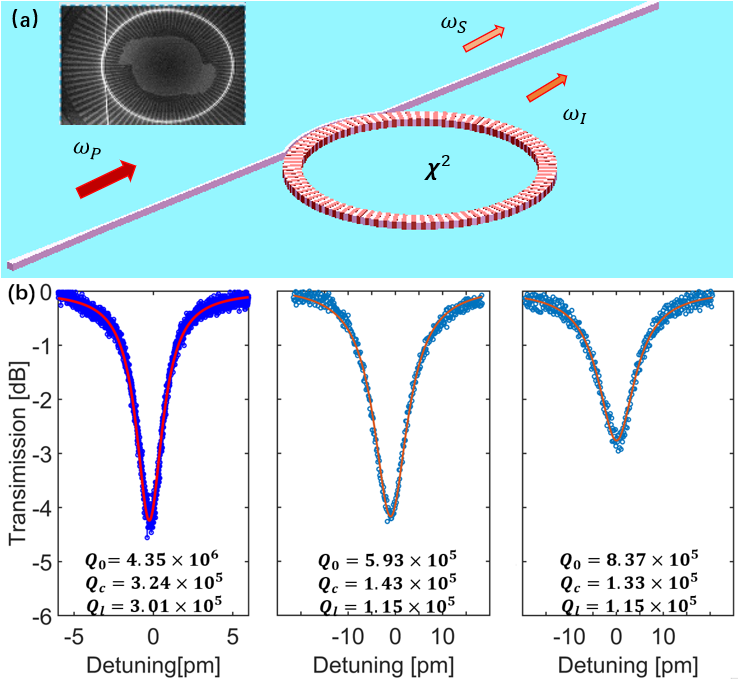}
\caption{\label{fig:1} (a): Schematic of the Z-cut periodical poling microring resonator, where the pump 
($\omega_P$) couples into the microring and generates signal ($\omega_S$) and idler ($\omega_I$). A pulley coupler is designed for overcoupling all light waves for high photon-extraction efficiency. Inset shows an SEM image of the microring with the pulley waveguide. Figure 1 (b) plots the typical spectra of interacting TM$_{00}$ cavity modes at (i) 776.96 nm, (ii) 1551.85 nm, and (iii) 1555.93 nm}
\end{figure}

\textbf{Device Calibration and Experiment Setup}. 
Figure 1 gives device details of the on-chip multiphoton source. As shown in Fig.~\ref{fig: Device Calibration}.(a), it is a perodically poled microring cavity fabricated on a Z-cut LNOI wafer (by NANOLN Inc.), with a 600-nm thick lithium niobate thin film bonded onto a 2-$\mu$m silicon dioxide layer above a silicon substrate. Utilizing our standard fabrication method \cite{chen2021photon}, a top width of 1.6$\mu$m and a radius of 80$\mu$m perodically poled lithium niobate (PPLN) microring is etched with a pulley bus wavguide as the coupler. The loaded quality factor ($Q_l$) is measured for each mode, and the coupling ($Q_c$) and intrinsic ($Q_0$) factors are each calculated by fitting the resonance spectra; see result in Fig.~\ref{fig: Device Calibration}.(b). The chip is fiber coupled, with the fiber-chip-fiber coupling losses measured to be 9.2 ± 0.2 dB at 1553.93 nm and 11.5 ± 0.3 dB at 776.96 nm, respectively. The overall optical nonlinearity is characterized by second harmonic generation (SHG), similarly to our previous measurement \cite{chen2019ultra}. With an on-chip pump power $P_p$ of 4.78$\mu$W, $P_\mathrm{SH}=75$ nW of second harmonic light is coupled out into the bus waveguide. The SHG efficiency is thus $\eta_\mathrm{SHG} =P_\mathrm{SH}/P^2_{p}=0.33\%/\mu W$, thus supporting highly efficient SPDC using only microwatt pumping. 

In a single-mode cavity, the effective Hamiltonian describing quasi-phase matched, non-degenerate spontaneous parametric downconversion can be written as follows:
\begin{equation}
\hat{H}_\textbf{eff} = \hbar\mathnormal{g}(\hat{a}_s\hat{a}_i\hat{b}_p^\dagger+\hat{a}_s^\dagger\hat{a}_i^\dagger\hat{b}_p),
\label{eq1}
\end{equation}
where $\{$$\hat{a}_s$, $\hat{a}_i$, and $\hat{a}_p$$\}$ each denotes the annihilation operator for the signal, idler, and pump photons, and $\mathnormal{g}$ is the nonlinear coupling coefficient between the pump and photon pairs. By periodic poling, the current lithium niobate micro-ring resonator can achieve phase matching while attaining the largest overlap between the fundamental quasi-transverse magnetic (quasi-TM) cavity modes in the infrared bands for the signal and idler photons and the visible band for the pump. Meanwhile, it provides the access to the largest $\chi^{(2)}$ nonlinear tensor $d_{33 }$ of lithium niobate. All contribute to a large effective nonlinear coupling coefficient $g$, which is given by \cite{chen2021photon} 
\begin{equation}
{g}=\sqrt{\frac{\hbar\omega_p\omega_s\omega_i}{2\epsilon_0\epsilon_p\epsilon_s\epsilon_i}} \frac{\frac{2}{\pi}d_\mathrm{eff} \zeta}{\sqrt{V_\mathrm{eff}}},
\label{eq2}
\end{equation}
where $\omega_j$ is the angular frequency, with $j$=$p$,$s$, and $i$ indicates the pump, signal, and idler modes, respectively. $\epsilon_0$ is the vacuum permittivity. $\{\epsilon_j\}$ are the relative permittivities. $d_\mathrm{eff}$ is the effective nonlinear susceptibility with the quasi phase matching discount. $\zeta$ is the mode-overlapping factor. $V_\mathrm{eff}$ is the effective mode volume. For the current microring device, the calculated single-photon coupling strength $g$ is 2.98 MHz.  

\begin{figure}[ht]
  \label{fig: Experiment Setup}
  \centering
\includegraphics[width=3.2in]{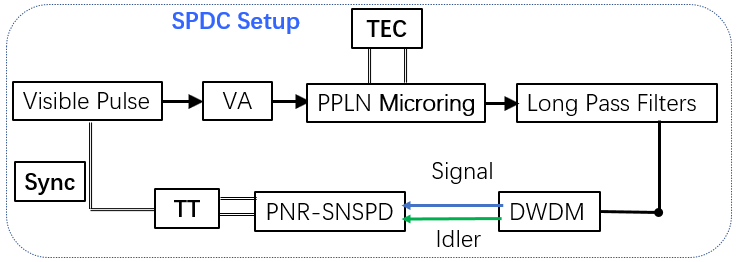}
\caption{\label{fig:2} Experiment setup. VA, Variable Attenuator; DWDM, Dense wavelength-division multiplexing; TT, Time Tagger unit; Sync, synchronise cable.}
\end{figure}

\begin{figure*}[ht]
\label{fig: Photon number statisitcs}
  \centering
\includegraphics[width=7.5in, height=3in]{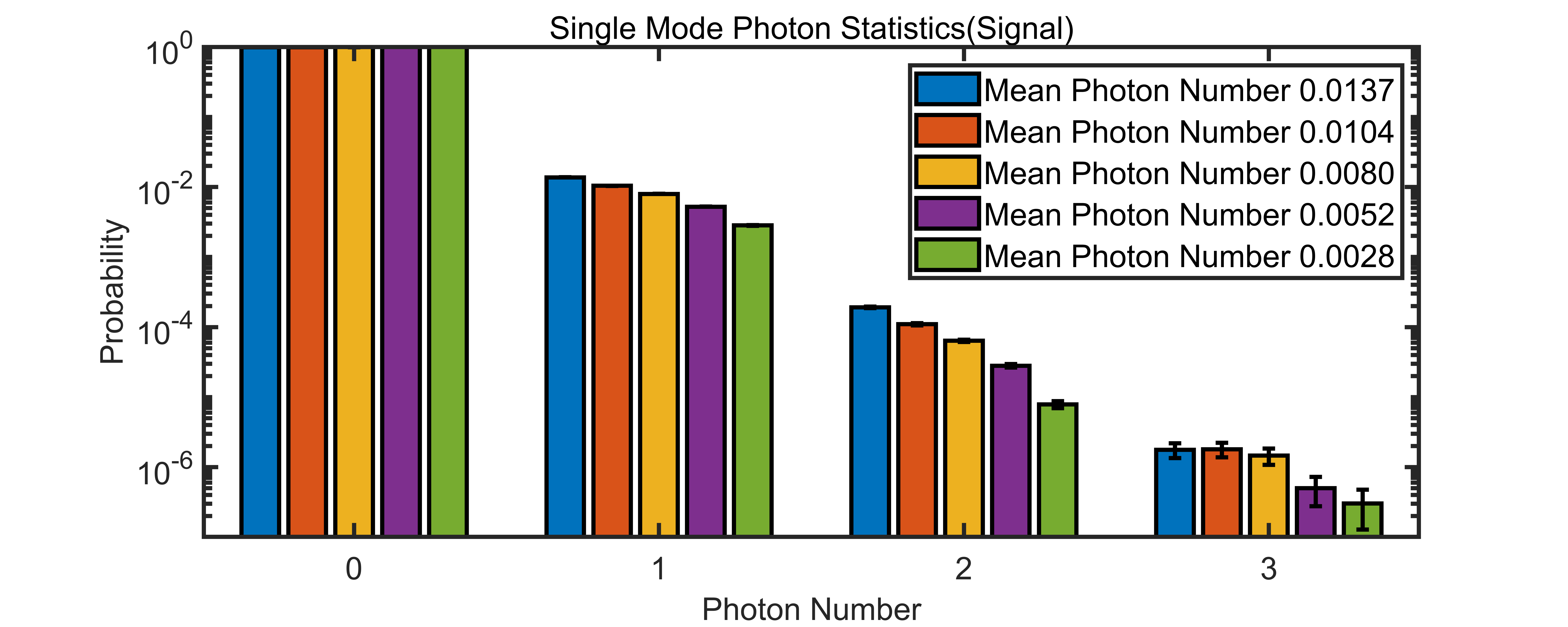}
\caption{\label{fig:3} Photon number statistics for different mean photon numbers.}
\end{figure*}

Figure 2 illustrates the experiment setup for generating and detecting photons. A visible pulse train with a pulse duration of $\tau$=300 ps and a repetition rate of 10MHz is created with a bulk SHG system made of a periodic-poled lithium niobate waveguide, to match the cavity lifetime and ensure single-mode operations; see details in Appendix A. Its power is varied by using a visible fiber attenuator (OZ OPTICS), and its polarization is controlled using fiber polarization controllers (FPCs). The output is fed into the microring cavity to excite the quasi-TM visible mode at 776.96 nm with a bandwidth of 1.14 GHz. There, signal and idler photons are created through SPDC into 1551.85 nm and 1555.93 nm quasi-TM modes, respectively, each in bandwidth of 1.68 GHz. The SPDC efficiency is tracked and maximized by temperature tuning using a temperature electronic controller (TEC). Subsequently, the generated photon pairs are filtered using an inline long-pass filter featuring an 80 dB extinction ratio and a 0.5 dB insertion loss (IL) to eliminate the pump power while transmitting the generated photon pairs.

In order to separate the signal and idler photons, cascaded dense wavelength division multiplexing (DWDM) filters with a full width at half maximum (FWHM) transmission bandwidth of 1.6 nm are employed, resulting in a transmission loss of approximately 0.3 dB. A pair of FPCs are then utilized to independently prepare the signal and idler photons in the best polarization states to be detected by the SNSPDs with the maximum detection efficiency. The two-channel PNR-SNSPDs (ID281, ID Quantique) feature a dark count rate of 50 to 100 Hz and detection efficiencies of 70\% (corresponding to 1.55~dB loss) and 82\% (0.86~dB loss), respectively. The detector outputs are fed to a synchronized time-tagging unit (Swabian Instrument). Accounting for all insertion losses, chip-fiber coupling losses, and finite detection efficiencies, the signal and idler channels experience a total loss of $\eta_\mathrm{S}$=7.55 dB and $\eta_\mathrm{I}$=6.76 dB, respectively. 


  \begin{figure}[ht]
  \label{fig: Fitting}
  \centering
\includegraphics[width=3.5in, height=1.5in]{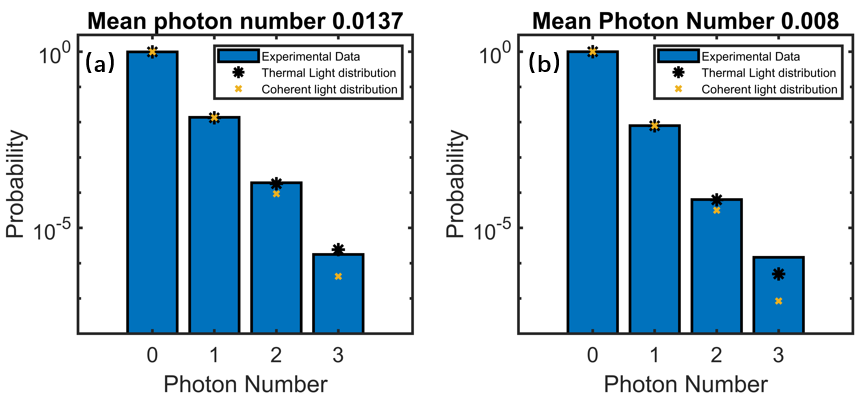}
\caption{\label{fig:4} (a) Detected photon distribution,  thermal light and coherent light fitting at a mean photon number of approximately 0.0137. (b) Detected photon distribution, thermal light and coherent light fitting at a mean photon number of approximately 0.008.}
\end{figure}

\textbf{Photon Number Statistics}. Upon carefully calibrating the chip device and SNSPDs, we proceed to measure the photon number statistics of the signal and idler photons while varying the input pump power. For the signal channel, the measurement results are shown in Fig.~\ref{fig:3}, where the normalized probabilities of detecting 0, 1, 2, and 3 photons are plotted along with error bar (assuming shot noise) under various mean photon number. As shown, the overall photon number distribution follows thermal distribution, as expected. As the mean photon number increases, the relative probabilities of multiple photon events increase. For example, when the mean photon number is 0.0028, the normalized probabilities for one, two, and three photons are $2.80\times10^{-3}$, $7.85\times10^{-6}$, and $3.03\times10^{-7}$, respectively. As we increase it to 0.0137, they each become $1.37\times10^{-2}$, $1.91\times10^{-4}$, and $1.77\times10^{-6}$. Because the SPDC saturated regime of multi-photon starts at mean photon number of 0.0137, there will be no obvious increasing at three-photon case.  In the figure, the error bars for three-photon events are higher because of much less detection events so that the Poissonian noise is more pronounced. 

To further show that our SPDC source indeed operates in the single-mode range, in Fig.~\ref{fig:4} we compare the measurement results with ideal thermal light distribution (TLD) in a single mode for two mean photon number cases: 0.0137 and 0.008. The TLD follows $P(n)=\bar{n}^n/(1+\bar{n})^{n+1}$, where $n$ and $\bar{n}$ denote the photon number and their mean, respectively. As seen, in both cases the measurement results agree well with TLD for the one and two photon cases. Compared with the coherent light distribution, there is a clearly deviation. For the three photon case, there is noticeable discrepancy, which can primarily be ascribed to the threshold sensitivity encountered in higher photon situations for the present SNSPD system. These results verify that our SPDC photons are in single modes, as desirable for many quantum information and quantum computing processes.


\textbf{Photon correlation}. Next, we characterize the one-photon and two-photon pair generation, by measuring their rates in each individual channel and jointly over paired SPDC channels. Specifically, we record the events of detecting one and two photons in the signal channel, with rates $N_{S}$ and $N_{SS}$, respectively, and in the idler channel with $N_{I}$ and $N_{II}$. Simultaneously, we record the one-photon coincident events where there is one photon detected in each channel, with rate $N_{SI}$, as well as two-photon coincident events for two photons per channel with rate $N_{SSII}$. 

\begin{figure}[ht]
  \label{fig: PGR}
  \centering
\includegraphics[width=3.5in, height=1.5in]{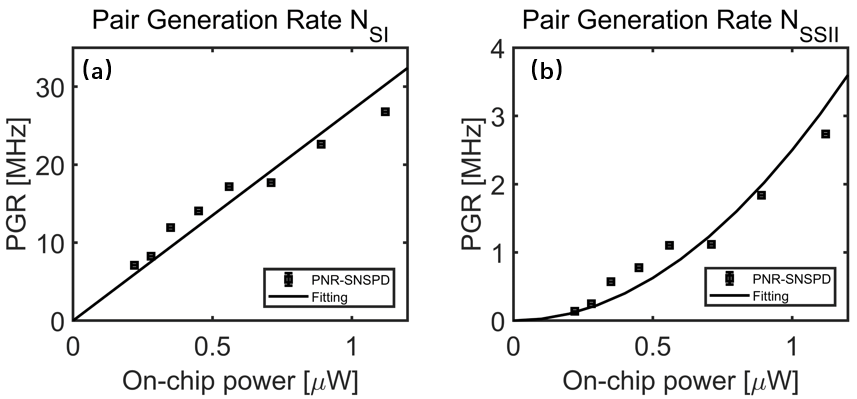}
\caption{\label{fig: 5} On-chip generation rates for one-photon pairs (a) and two-photon pairs (b), respectively, along with their curve fitting results.}
\end{figure}

From these rates, the on-chip generation rate for the one-photon pairs is estimated to the first order as $P_{SI}=N_{S}N_{I}/N_{SI}$. The results are plotted as a function of the on-chip SPDC pump power in Fig.~\ref{fig: 5}.(a). As shown, $P_{SI}$ increases linearly with the power, as expected. Only 220 nW power is needed to create 7 million pairs per second. By linear regression, the brightness, defined as pair generation per unit pump power, is obtained as the slope of the fitting curve as 27 MHz/${\mu}$W, which is among the highest across all SPDC sources in various materials. The detection rate corresponds to ten times higher than our previous result \cite{ma2020ultrabright}, which is ascribed to the higher efficiencies in both photon pair generation and detection.  

Similarly, for the two-photon pairs (i.e., two signal photons and two idler photons generated simultaneously in pairs), the on-chip rate under first order approximation is $P_\mathrm{SSII}=N_\mathrm{SS}N_\mathrm{II}/N_\mathrm{SSII}$. The results as a function of the on-chip pump power are plotted in Fig.~\ref{fig: 5}.(b). In contrast to the one-photon pair case, here the rate increases quadratically over the power, because the underline process is of the second order in SPDC. At 220 nW pumping, the two-photon pair on-chip rate is 8.6($10^4$), and increase to 9.5($10^6$) at 1.12 $\mu$W.

\begin{figure}[ht]
  \label{fig: Pnm}
  \centering
\includegraphics[width=3.5in, height=2in]{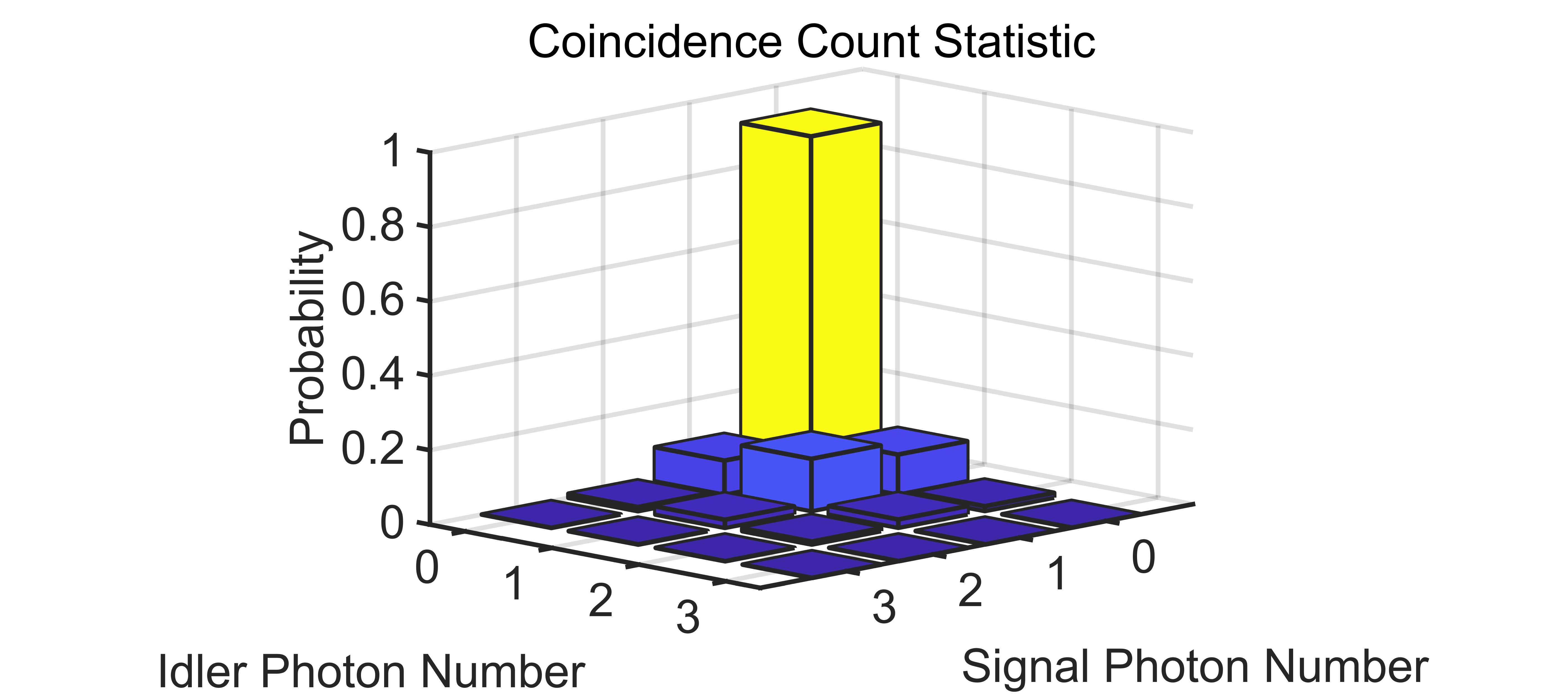}
\caption{\label{fig: 6} Coincidence photon probability at average pump power around 1.1$\mu$W.}
\end{figure}

The above results are from simple calculations under the first order approximation. To further characterize the mutliphoton correlation, we count the joint events of mixed photon numbers and use loss inversion to calculate the inferred joint states of photon numbers \cite{harder2016single}. The results for 0.137 mean photon number on chip are shown in Fig.~\ref{fig: 6}, where we neglect the contributions from detector dark counts and ambient photons (about 100 Hz). As seen, while the photon numbers in the signal and idler channels are correlated, the correlation is not strong. This is mainly due to the high total loss of each channel (7.55 dB and 6.76 dB) and the low coincidence events of multiphoton states, because of which the loss inversion calculation is not very accurate. 

\begin{table*}
\begin{center}\caption{\label{demo-table}CAR measured at average power around 1.1$\mu$W}
\scalebox{1.2}{
\begin{tabular}{r|llll}
\multicolumn{1}{r}{}
& \multicolumn{1}{l}{~~~~~~~~~~~S(0)}
& \multicolumn{1}{l}{~~~~~~~~~~~S(1)}
& \multicolumn{1}{l}{~~~~~~~~~~~S(2)}
& \multicolumn{1}{l}{~~~~S(3)} \\ \cline{2-5}
I(0) &$1.17\times 10^9$:$1.16\times10^9$ & $1.34\times10^7$:$1.51\times10^7$ & $1.54\times10^5$:$1.93\times10^5$ &296:373\\
I(1) & $1.44\times10^7$:$1.61\times10^7$ & $1.84\times10^6$:$2.10\times10^5$  & $3.88\times10^4$:$2.73\times10^3$ &~~77:5 \\
I(2) &$1.80\times10^5$:$2.24\times10^5$       & $4.38\times10^4$:$2.93\times10^3$      & $3.52\times10^3$:35     & ~~~8:0   \\
I(3) &$1.52\times10^3$:$1.94\times10^4$          & ~~~~~~~~~412:22        & ~~~~~~~~~~~35:0       & ~~~3:0    \\
\end{tabular}}
\end{center}
\end{table*}
To get a better measurement, we exam the coincident detection of the various multiphoton states. The results for the same pump power as in Fig.~\ref{fig: 6} are given in Table I as the coincidence-to-accidental counting rates of one, two, and three photons in each channel. Here, the coincidence rates between S($n$) and I($m$) are for event occurrences of simultaneously detecting $n$ signal photons and $m$ idler photons in the same time slot (in this case each of 400 ps width). The accidental rates are for those events occuring in a different slot, set by 100 ns apart to avoid any correlation. As seen, the coincidence to accidental detection ratio is about 10 for single-photon pairs, and 100 for two-photon pairs, which shows high correlation. Over our total acquistion period of 120 seconds, we record 3 coincidence of three photon pairs, but no accidental event. Interestingly, in the Figure, the coincident rates are not maximized at diagonal. For example, the coincident detection of one signal photon and two idler photons is more likely than that of two signal and two idler photons. This is because although signal and idler photons are created on chip with strong photon number correlation, the total loss is about 7 dB per channel so that only a fraction of them can be detected thus blurring the correlation.       

From Table I, the mutual correlation function can be calculated $g^{(n,m)}=\langle\hat{a}^{\dagger n}\hat{a}^{ n}\hat{b}^{\dagger m}\hat{b}^{m}\rangle/{\langle\hat{a}^\dagger\hat{a}\rangle}^n{\langle\hat{b}^\dagger\hat{b}\rangle}^m$. To satisfy the non-classical criteria \cite{avenhaus2010accessing,vogel2008nonclassical}, the following condition must be met: $\gamma=g^{(1,2)}/\sqrt{g^{(2,2)}g^{(0,2)}}>1$. From the pump power ranging from 220 nW to 1.12 $\mu$W, we have calculated $\gamma$ to between 1.3 and 1.6, indicating good quantum correlation.

  \begin{figure}[ht]
  \label{fig: Heralded g2}
  \centering
\includegraphics[width=3.5in, height=2.5in]{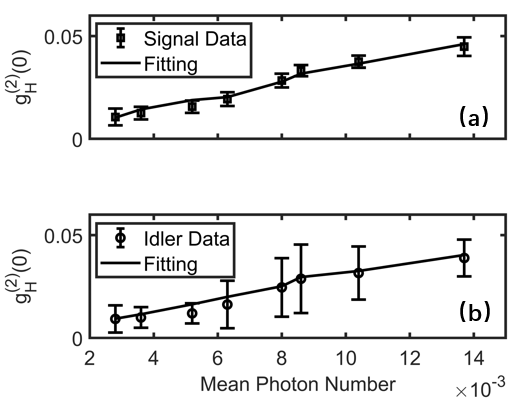}
\caption{\label{fig: 7} Heralded $g_H^{(2)}{(0)}$ for signal (a) and idler photons (b). }
\end{figure}

\begin{table*} 
\caption{Mode Purity in Various Photon Sources}
\begin{tabular}{c | c | c | c | c | c | c} 
 \hline
 Reference & Material Structure & Quality Factor & Pulse Width & $g^{(2)}$ & $K$ & Purity \\ [0.5ex] 
 \hline\hline

Eckstein\cite{eckstein2011highly} & PPKTP\footnote{Periodically Poled Potassium Titanyl Phosphate} Waveguide & N/A & 1ps & 1.95 & 1.05 & 95\%\\
 \hline
 Harder\cite{harder2016single} & PPKTP Waveguide & N/A  & 1ps & 1.89 & 1.12 & 89\%  \\
 \hline
Stasi\cite{stasi2022enhanced} & PPKTP Waveguide & N/A  & 1ps & 1.99 & 1.01 & 99\% \\
 \hline 
 Vaidya\cite{vaidya2020broadband} & $Si_3N_4$  $\mu$-ring & $8(10^5)$ & 1.5ns & 1.95 & 1.05 & 95\%  \\ 
 \hline
 This work & PPLN $\mu$-ring & $1.15(10^5)$ & 300ps & 1.99 & 1.01 & 99\% \\
 \hline\hline
\end{tabular}
\end{table*}

We next study the prospective use of this source for heralded single photon generation.  Figure~\ref{fig: 7} plots the heralded photon correlation for both channels under various pump power. In contrast to standard Hanbury Brown and Twiss effect (HBT) measurement using a beamsplitter, here we utilize the collected multiphoton statistics directly by the PNR-SNSPDs. In this case, the second-order correlation function at $\tau$=0 without heralding, denoted as $g^{(2)}(0)$, is given by $g^{(2)}(0) = \sum n(n-1)P(n)/(\sum nP(n))^2$. The results are around 1.99 to 2.25 (see given in the Appendix B), verifying the thermal statistics of each SPDC channel under the single-mode condition. In the heralding case, on the other hand, the same statistics is taken only when there is one photon clicking event in the paired channel. In this case, the correlation becomes $g^{(2)}_H(0) =\sum n(n-1)P(n|1)/(\sum n P(n|1))^2$, where $P(n|m)=P(n,m)/P(m)$ is the conditional probability of detecting $n$ photons in one channel upon detecting $m$ photons in the other, computed from the joint detection probability of $m$ and $n$ photons in the two channels and that of a single one. With the coincidence counts from two PNR-SNSPDs, we can easily compute $P(n|1)$ for both signal and idler channels. As seem in the figure, for both channels, $g^{(2)}_H(0)$ is about 0.01 when the mean photon numbers are $0.003$ per pulse and increases to approaching 0.05 as the mean photon numbers increase to $0.014$.

Finally, we compare the time-frequency mode purity obtained here with competing sources. The results are summarized in Table II. In waveguides, it typically requires to use picosecond pump pulses so as to match the optical filters for the generated photons, to obtain single modes. In comparison, those based on resonators, such as the present microrings, the  pump pulses can be much longer, ranging from a few hundred picoseconds to nanoseconds in order to match with the cavity's lifetime for single modes. In this device, the effective mode number $K=1/[g^{(2)}(0)-1]$ is at 1.01, which is very close to the ideal case with $K=1$ \cite{harder2016single}. This represents a mode purity of $1/K=99 \pm 4.9\%$, indicating an optimal condition for single mode photon production, as desirable for many applications.

In conclusion, we have demonstrated photon statistics with a two-channel PNR-SNSPD system, characterizing single-photon and multiphoton pair generation. Utilizing an ideally quasi phase matched lithium niobate microring in Z cut, we have scored a ten-fold enhancement in the SPDC generation rate of single-photon pairs \cite{ma2020ultrabright}. We measured joint photon probabilities of multiphoton states up to three photons in a channel. Also, we have performed coincident to accidental photon detection for multiphoton states using time-delayed measurement, for the first time. Our results highlight a SPDC source for multiphoton entanglement with both high efficiency and mode purity, as needed for many quantum information processing applications with multiphoton states. This work paves the way for the development of advanced quantum photonic devices and systems with good performance and versatility.

\begin{figure*}[ht]
  \centering
\includegraphics[width=6.5in]{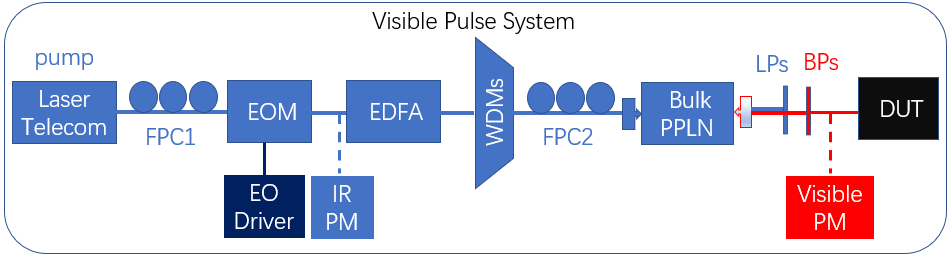}
\caption{\label{fig:epsart} Setup for generate the SPDC pump. Blue and red lines depict the telecom light path and visible path, respectively. FPC, fiber-polarization controller;  EOM, Electro-Optic modulator; PM, power meter; EDFA, Erbium-Doped Fiber Amplifier;  WDM, wavelength division multiplexing module; LP, Low pass filter; BP, bandpass filter; 
DUT, device under test. }
\end{figure*}

\begin{acknowledgments}
The research was supported in part by the Office of Naval Research (Award No. N00014-21-1-2898) and by ACC-New Jersey (Contract No. W15QKN-18-D-0040). Device fabrication was performed at ASRC, CUNY.

\end{acknowledgments}

\appendix

\section{SPDC Pump Generation}

\begin{figure}[ht]
  \centering
\includegraphics[width=3.5in, height=2.5in]{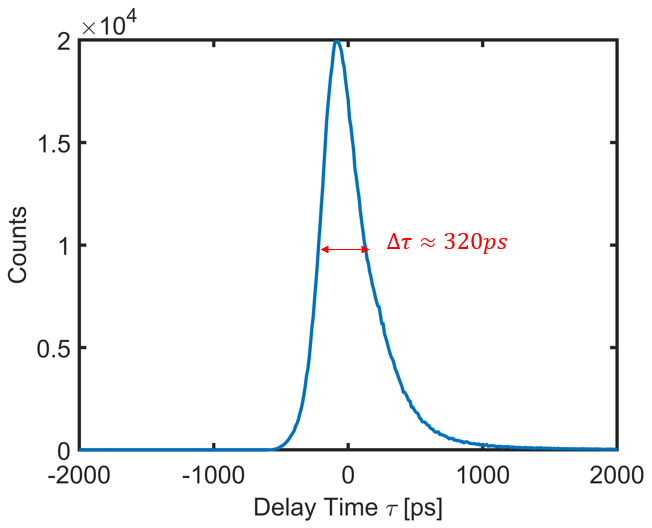}
\caption{\label{fig:epsart} Pulse width measurement}
\end{figure}

To create the SPDC pump pulses in the visible band, a single-channel picosecond EOM (electro-optical modulator) driver (Highland Technology T130, 250ps-30ns pulse width, 0-50MHz pulse rate) supplies Radiofrequency pulse(10MHz) to the EOM. An IR power meter monitors the EOM output. An erbium-doped fiber amplifier (EDFA) in the telecom C band further amplifies the weak signal, followed by two DWDM filters to clean the beam. The resulted signal then couples into a bulk PPLN crystal to create the visible pulsed light as the SPDC pump. Two low-pass-filters(IL $\sim$ 0.5 dB; extinction ratio, ER $\sim$ 50 dB) and narrow-band-pass-filters(Alluxa, 3 nm, IL$\sim$ 1 dB, ER $>$ 120 dB) reject the pump signal and passing largest light at 776.96nm.

In Figure 9, we measure the cross-correlation between photons created by the signal cavity and the synchronized electronic pulse from EO Driver by using a time tagger. The full width at half maximum is around 320 ps.  Due to the EO Driver jitter(10 ps) and PNR-SNSPDs jitter(54 ps), it is slightly wider than the electronic pulse(300 ps).

\section{Signal and Idler $g^{(2)}{(0)}$ Measurement}


Figure 10(a) and 3(b) plot the photon correlation measurement of the signal and idler channel using two PNR-SNSPDs, before heralding. Here the second-order correlation function is calculated from the PNR-SNSPD results as $g_\mathrm{unc}^{(2)}(0) = \sum n(n-1)P(n)/(\sum nP(n))^2$. As seen, for both channels $g^{(2)}_\mathrm{unc}(0) \approx 2$ for each channel at different mean photon number.
\begin{figure}[ht]
  \centering
\includegraphics[width=3.5in, height=2.5in]{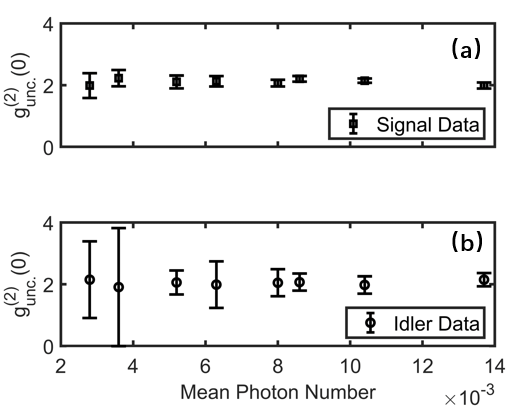}
\caption{\label{fig:epsart} (a) and (b): Unheralded two-photon correlation in signal and idler channels.}
\end{figure}


\bibliography{apssamp}

\end{document}